\documentclass{pasj00}
\DeclareAbbreviation\AAHam{Astron. Abh. Hamburg. Sternw.}
\DeclareAbbreviation\AARv{Astron. Astrophys. Rev.}
\DeclareAbbreviation\an{Astron. Nachr.}
\DeclareAbbreviation\AcA{Acta Astron.}
\DeclareAbbreviation\Afz{Astrofizika}
\DeclareAbbreviation\AnTok{Tokyo Astron. Obs. Annals, Sec. Ser.}
\DeclareAbbreviation\Ap{Astrophysics}
\DeclareAbbreviation\ARep{Astron. Rep.}
\DeclareAbbreviation\ATel{Astronomer's Telegram}
\DeclareAbbreviation\ATsir{Astron. Tsirk.}
\DeclareAbbreviation\AcApS{Acta Astrophys. Sinica}
\DeclareAbbreviation\AstL{Astron. Letters}
\DeclareAbbreviation\BaltA{Baltic Astron.}
\DeclareAbbreviation\BASI{Bull. Astron. Soc. India}
\DeclareAbbreviation\BeSN{Be Star Newsletter}
\DeclareAbbreviation\GCN{GCN}
\DeclareAbbreviation\ibvs{Inf. Bull. Variable Stars}
\DeclareAbbreviation\JAD{J. Astron. Data}
\DeclareAbbreviation\JAVSO{J. American Assoc. Variable Star Obs.}
\DeclareAbbreviation\JBAA{J. British Astron. Assoc.}
\DeclareAbbreviation\LowOB{Lowell Obs. Bull.}
\DeclareAbbreviation\MitVS{Mitteil. Ver\"{a}nderl. Sterne}
\DeclareAbbreviation\MmSAI{Mem. Soc. Astron. Ita.}
\DeclareAbbreviation\Msngr{Messenger}
\DeclareAbbreviation\NewA{New Astron.}
\DeclareAbbreviation\NewAR{New Astron. Rev.}
\DeclareAbbreviation\OAP{Odessa Astron. Publ.}
\DeclareAbbreviation\Obs{Observatory}
\DeclareAbbreviation\PASA{Publ. Astron. Soc. Australia}
\DeclareAbbreviation\PAZh{Pis'ma AZh}
\DeclareAbbreviation\PhR{Phys. Rep.}
\DeclareAbbreviation\PVSS{Publ. Variable Stars Sect. R. Astron. Soc. New Zealand
}
\DeclareAbbreviation\PZ{Perem. Zvezdy}
\DeclareAbbreviation\PZP{Perem. Zvezdy Pril.}
\DeclareAbbreviation\QJRAS{QJRAS}
\DeclareAbbreviation\RMxAA{Rev. Mexicana Astron. Astrof.}
\DeclareAbbreviation\RvMA{Reviews of Modern Astron.}
\DeclareAbbreviation\Sci{Science}
\DeclareAbbreviation\SvA{Soviet Astronomy}
\DeclareAbbreviation\SvAL{Soviet Astronomy Letters}
\DeclareAbbreviation\VeSon{Ver\"{o}ff. Sternw. Sonneberg}
\DeclareAbbreviation\VSOLJBul{VSOLJ Variable Star Bull.}
\DeclareAbbreviation\yCat{VizieR Online Data Catalog}
\DeclareAbbreviation\ZA{Z. Astrophys.}

\def\ASPConf#1#2{ASP Conf. Ser. #1, #2}

\def\PublisherASP{San Francisco: ASP}

\begin{document}
\SetRunningHead{M. Uemura, et al., }{WZ Sge-type dwarf nova, SDSS J102146.44$+$234926.3}
\Received{2007/10/26}%{yyyy/mm/dd}
\Accepted{2007/11/18}%{yyyy/mm/dd}

\title{Discovery of a WZ Sge-Type Dwarf Nova,
  SDSS~J102146.44$+$234926.3: Unprecedented Infrared Activity during a
  Rebrightening Phase}

%%% begin:list of authors
\author{
Makoto \textsc{Uemura}\altaffilmark{1},
Akira \textsc{Arai}\altaffilmark{2},
Tom \textsc{Krajci}\altaffilmark{3},
Elena \textsc{Pavlenko}\altaffilmark{4}
Sergei Yu. \textsc{Shugarov}\altaffilmark{5}\\
Nataly A. \textsc{Katysheva}\altaffilmark{5}
Vitalij P. \textsc{Goranskij}\altaffilmark{5}
Hiroyuki \textsc{Maehara}\altaffilmark{6},
Akira \textsc{Imada}\altaffilmark{6},\\
Taichi \textsc{Kato}\altaffilmark{6}, 
Daisaku \textsc{Nogami}\altaffilmark{7}, 
Kazuhiro \textsc{Nakajima}\altaffilmark{8},
Takashi \textsc{Ohsugi}\altaffilmark{1,2},\\
Takuya \textsc{Yamashita}\altaffilmark{1},
Koji S. \textsc{Kawabata}\altaffilmark{1},
Osamu \textsc{Nagae}\altaffilmark{2},
Shingo \textsc{Chiyonobu}\altaffilmark{2},\\
Yasushi \textsc{Fukazawa}\altaffilmark{2},
Tsunefumi \textsc{Mizuno}\altaffilmark{2},
Hideaki \textsc{Katagiri}\altaffilmark{2},
Hiromitsu \textsc{Takahashi}\altaffilmark{2},\\
Atsushi \textsc{Ueda}\altaffilmark{2},
Takehiro \textsc{Hayashi}\altaffilmark{9},
Kiichi \textsc{Okita}\altaffilmark{10},
Michitoshi \textsc{Yoshida}\altaffilmark{10},\\
Kenshi \textsc{Yanagisawa}\altaffilmark{10},
Shuji \textsc{Sato}\altaffilmark{11},
Masaru \textsc{Kino}\altaffilmark{11}, and
Kozo \textsc{Sadakane}\altaffilmark{12}}

\altaffiltext{1}{Astrophysical Science Center, Hiroshima University, Kagamiyama
1-3-1, \\Higashi-Hiroshima 739-8526}
\email{uemuram@hiroshima-u.ac.jp}
\altaffiltext{2}{Department of Physical Science, Hiroshima University,
Kagamiyama 1-3-1, \\Higashi-Hiroshima 739-8526}
\altaffiltext{3}{P.O. Box 1351, Cloudcroft, NM 88317, USA}
\altaffiltext{4}{Crimean Astrophysical Observatory, Crimea, Nauchnyj, Ukraine}
\altaffiltext{5}{Sternberg Astronomical Institute, Moscow, Russia}
\altaffiltext{6}{Department of Astronomy, Faculty of Science, Kyoto University, Sakyo-ku, Kyoto 606-8502}
\altaffiltext{7}{Hida Observatory, Kyoto University, Kamitakara, Gifu 506-1314}
\altaffiltext{8}{VSOLJ, 124 Isatotyo Teradani, Kumano, Mie 519-4673}
\altaffiltext{9}{Department of Education, Hiroshima University,
Kagamiyama 1-1-1, \\Higashi-Hiroshima 739-8524}
\altaffiltext{10}{Okayama Astrophysical Observatory, National
Astronomical Observatory of Japan, \\Kamogata Okayama 719-0232}
\altaffiltext{11}{Department of Physics, Nagoya University, Furo-cho,
Chikusa-ku, Nagoya 464-8602}
\altaffiltext{12}{Astronomical Institute, Osaka Kyoiku University,
Asahigaoka, Kashiwara, Osaka 582-8582}
%%% end:list of authors

%%% Please use the following style in case that sorting by 
%%% affiliation is impossible. 
%
% \author{%
%   D-Firstname \textsc{D-Familyname}\altaffilmark{1}
%   E-Firstname \textsc{E-Familyname}\altaffilmark{1,2}
%   and
%   F-Firstname \textsc{F-Familyname}\altaffilmark{2}}
% \altaffiltext{1}{Address of Institute}
% \email{ddddd@xxx.xxx.xx.xx}
% \email{eeeee@xxx.xxx.xx.xx}
% \altaffiltext{2}{Address of Institute}

%% `\KeyWords{}' always has to be placed before `\maketitle'.
 %Do NOT move this preamble from here!

\maketitle

\begin{abstract}
Several SU~UMa-type dwarf novae, in particular, WZ~Sge-type stars tend
to exhibit rebrightenings after superoutbursts.  The rebrightening
phenomenon is problematic for the disk instability theory of dwarf
novae since it requires a large amount of remnant matter in the disk
even after superoutbursts.  Here, we report our optical and infrared
observations during the first-ever outburst of a new dwarf nova, 
SDSS~J102146.44$+$234926.3.  During the outburst, we detected
superhumps with a period of $0.056281\pm 0.000015\,{\rm d}$, which is
typical for superhump periods in WZ~Sge stars.  In conjunction with
the appearance of a long-lived rebrightening, we conclude that the
object is a new member of WZ~Sge stars.  Our observations,
furthermore, revealed infrared behaviors for the first time in the
rebrightening phase of WZ~Sge stars.  We discovered prominent infrared
superhumps.  We calculate the color temperature of the infrared
superhump source to be 4600---6400~K.  These temperatures are too low
to be explained with a fully-ionized disk appearing during dwarf nova
outbursts.  We also found a $K_{\rm s}$-band excess over the hot disk
component.  These unprecedented infrared activities provide evidence
for the presence of mass reservoir at the outermost part of the
accretion disk.  We propose that a moderately high mass-accretion rate
at this infrared active region leads to the long-lived rebrightening
observed in SDSS~J102146.44$+$234926.3. 
\end{abstract}

\section{Introduction}

Dwarf novae are a sub-class of cataclysmic variables which contain a
Roche-lobe filling secondary star and an accreting white dwarf (for a
review, see \cite{war95book}).  Dwarf nova outbursts are characterized
with amplitudes of 2---5~mag and durations of a few days.  SU~UMa-type
stars are dwarf novae showing two types of outbursts, that is, normal
outbursts and superoutbursts which have longer durations ($\sim
10$---$20$~d) and amplitudes $\sim 1\,{\rm mag}$ larger than those of
normal outbursts.  Superoutbursts are characterized by the appearance
of periodic short-term modulations, called ``superhumps'', whose
period is slightly longer than the orbital period
(\cite{war95suuma}). 

Dwarf nova outbursts are caused by a sudden increase of mass-accretion
rate in an accretion disk.  The most widely accepted theory for SU~UMa
stars is the thermal---tidal instability model for the accretion disk
(\cite{osa89suuma}).  The thermal instability model considers two
kinds of thermally stable states of the disk, that is, a cool disk of
neutral hydrogen gas and a hot disk of ionized hydrogen gas,
corresponding to the quiescent and outburst states, respectively.
According to the disk instability model, the cool disk becomes
thermally unstable when the density of the accumulated gas reaches a
critical value.  The disk, then, experiences a state transition to the
hot disk, observed as an outburst (\cite{osa74DNmodel};
\cite{hos79diskinst}; \cite{mey81DNoutburst}; \cite{sma82alphadisk};
\cite{can82DNburst}).  The outburst leads to an expansion of the
disk.  In the case that the disk has an enough angular momentum, its
outer edge reaches the 3:1 resonance radius, which triggers the tidal
instability (\cite{whi88tidal}).  The disk becomes eccentric, as a
result, the strong tidal torque yields an additional dissipation of
the angular momentum in the disk, which leads to superoutbursts.
Because of the eccentricity of the disk, periodic modulations
associated with the orbital motion of the secondary star are expected.
They are actually observed as superhumps.  The superhump period is
longer than orbital periods due to the precession of the eccentricity
wave (\cite{pat05epsilon}). 

The disk instability model can explain several basic behaviors of
ordinary dwarf novae, for example, their frequency of outbursts and
two kinds of outbursts in SU~UMa stars (for a review, see
\cite{osa96review}).  This standard model is, however, required to be
modified in order to reproduce rebrightenings after superoutbursts
which are observed in several SU~UMa stars.  A peculiar subclass,
WZ~Sge-type dwarf novae, in particular, tend to exhibit various types
of rebrightenings (\cite{how95TOAD}; \cite{kat01hvvir}).  WZ~Sge-type
stars have shortest orbital periods ($P_{\rm orb}\sim 85\,{\rm min}$)
and longest outburst cycles ($\gtrsim 10\,{\rm yr}$) among dwarf
novae.  Based on characteristics of light curves, rebrightenings can
be divided into three types; i)~a single short rebrightening (RZ~Leo,
\cite{ish01rzleo}; ASAS~J002511$+$1217.2, \cite{tem06J0025}),
ii)~repetitive short rebrightenings, or sometimes called ``echo
outbursts'' (WZ~Sge, \cite{pat02wzsge}; EG~Cnc, \cite{pat98egcnc},
\cite{kat04egcnc}; SDSS~J080434.20$+$520349.2, \cite{pav07j0804}), and
iii) a long-lived plateau (AL~Com, \cite{nog97alcom},
\cite{ish02wzsgeletter}; CG~CMa, \cite{kat99cgcma};
TSS~J022216.4$+$412259.9, \cite{ima06j0222}; V2176~Cyg,
\cite{nov01v2176cyg}).  Additionally to WZ~Sge stars, several
short-period SU~UMa stars also exhibit single or a few short
rebrightenings (\cite{bab00v1028cyg}; \cite{uem02j2329};
\cite{ima06j0137}).  On the other hand, several WZ~Sge stars show no
sign of rebrightenings (HV Vir, \cite{kat01hvvir}).

The mechanism of rebrightenings and the origin of their diversity are
poorly understood.  In a simple picture based on the disk instability
model, most of the mass and angular momentum are lost from the disk by  
the end of superoutbursts (\cite{osa96review}).  The long-lived 
rebrightening phase is problematic because it apparently requires an
unknown delayed mass-accretion even after the termination of
superoutbursts when the disk should nearly be empty
(\cite{kat04egcnc}).  This discrepancy can be reconciled if a
mass-transfer rate from the secondary is enhanced due to the
irradiation by the disk (\cite{war98CVreviewWyo};
\cite{ham00DNirradiation}; \cite{pat02wzsge}).  This scenario,
however, has a difficulty to reproduce sudden onset and cessation of
the rebrightening phase (\cite{pat02wzsge}; \cite{osa03DNoutburst};
\cite{osa04emt}).  \citet{kat98super} propose an alternative scenario
that a considerable amount of matter left over the 3:1 resonance
radius is responsible for the delayed accretion.  The remnants at the
outermost region is expected to work as a mass reservoir, leading to 
the additional mass accretion during the rebrightening phase
(\cite{hel01eruma}; \cite{osa01egcnc}; \cite{kat04egcnc}). 

The behavior of the outermost region of the accretion disk, hence,
definitely provides crucial clues for the rebrightening phenomenon.
Infrared (IR) observations can be a strong tool to detect variations
from a low temperature region at the outermost accretion disk.  Such
IR observations during the rebrightening phase have, however, been
difficult because of low frequencies of outbursts of WZ~Sge stars. 

Here, we report results of our observation of a new variable star
which was a promising candidate for WZ~Sge stars.  The discovery of
its outburst was reported by E.~J.~Christensen in Nov. 2006
(\cite{chr06cbet746}).  There is a quiescent counterpart, 
SDSS J102146.44$+$234926.3 (hereafter, J1021) in the Sloan Digital Sky 
Survey (SDSS) DR5 database with magnitudes of $g=20.74$ and $r=20.63$
(\cite{SDSSDR5}).  We conducted an international campaign for J1021,
in which we succeeded in performing IR time-series observations during 
a rebrightening phase for the first time.  In the next section, we
describe our photometric observations.  In section~3, results of our
observations are presented.  In section~4, we discuss the results and
the nature of the rebrightening phenomenon.  In the final section, we
summarize our findings.

\section{Observations}

\begin{figure}
  \begin{center}
    \FigureFile(80mm,80mm){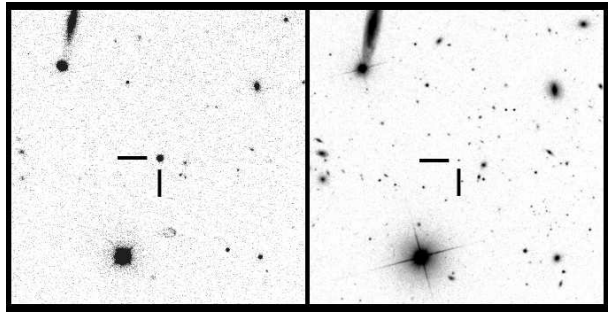}
    %%% \FigureFile(width,height){filename}
  \end{center}
  \caption{Optical images of the field of J1021 in SDSS DR5 (right
 panel; \cite{SDSSDR5}) and observed with the KANATA 1.5-m telescope
 (left panel).  The field of view is approximately
 \timeform{5'}$\times$\timeform{5'}.  SDSS~J1021 is the object marked
 with the black bars.}\label{fig:img} 
\end{figure}

\begin{table*}
  \caption{Observation log.}\label{tab:log}
  \begin{center}
    \begin{tabular}{cccp{2cm}p{3cm}}
     \hline \hline
     Site & Telescope & Camera & Filter & Date ($+$JD2454000) \\
     \hline
     Higashi-Hiroshima Astr. Obs. & KANATA 1.5-m & TRISPEC & $V$, $J$, $K_{\rm
     s}$, none (optical channel) & 70, 71, 74, 75, 76, 85, 88,
     89, 91, 92, 100, 115, 123, 125, 126, 127, 130 \\
     Cloudcroft & 28-cm & SBIG ST-7 & none & 60, 61, 62\\
     Crimean Laboratory of SAI & 50- and 60-cm & AP 47p & $V$, $R_{\rm
       c}$ & 62, 63, 64, 65, 66, 69, 72, 73, 75, 76, 77, 79, 80, 81,
     82, 84, 86, 90, 91\\ 
     Crimean Astr. Obs. (CrAO) & 2.6-m & FLI 1001E & $R_{\rm c}$ & 91\\
     Saitama & 25-cm & SBIG ST-7XME & clear & 61, 65, 71, 75\\
     Kyoto Univ. & 40-cm & SBIG ST-9 & none & 73, 74\\
     Mie & 25-cm & MUTOH CV04 & none & 73, 74, 75\\
     \hline
    \end{tabular}
  \end{center}
\end{table*}

Simultaneous optical and IR observations were performed with TRISPEC
attached to the KANATA 1.5-m telescope at Higashi-Hiroshima
Observatory.  TRISPEC is a simultaneous imager and spectrograph with
polarimetry covering both optical and near-infrared wavelengths 
(\cite{TRISPEC}).  We used the imaging mode of TRISPEC with $V$, $J$,
and $K_{\rm s}$ filters.  Effective exposure times for each frame were
63, 60, and 54~s in $V$, $J$, and $K_{\rm s}$ bands, respectively.
An example of the images is shown in the left panel of
figure~\ref{fig:img}.   The figure also includes the image of the same
field in the SDSS DR5 database in the right panel for comparison
(\cite{SDSSDR5}).  J1021 is the object marked with the black bars. 

After making dark-subtracted and flat-fielded images, we measured
magnitudes of J1021 and comparison stars using a Java-based aperture
photometry package.  The $V$, $J$, and $K_{\rm s}$ magnitudes of J1021
were calculated from those of the comparison star located at
R.A.$=$\timeform{10h21m53.33s}, Dec.$=$\timeform{+23D50'55.9''} 
($V=12.84$, $J=12.22$, $K_{\rm s}=12.04$).  We quote the $V$-band
magnitude of the comparison star listed in the Henden's
sequence\footnote{ftp://ftp.aavso.org/public/calib/varleo06.dat} and
$J$ and $K_{\rm s}$-band magnitudes from the 2MASS catalog
(\cite{2MASS}).  We checked the constancy of the comparison star using 
neighbor stars and found that it exhibited no significant variations
over 0.02, 0.03, and 0.06~mag in $V$, $J$, and $K_{\rm s}$-bands,
respectively.  For $V$- and $J$-band observations, we obtained
time-series data which allowed us to study short-term variations in a
night.  For $K_{\rm s}$-band observations, we only obtained averaged
data for each night.

We also performed optical photometric observations with other
telescopes during the period of JD~2454060---2454130 at 6 sites, that
is, Cloudcroft, Crimea, Saitama, Kyoto, and Mie.  Details of their
observational equipment is summarized in table~\ref{tab:log}.  The
magnitude systems with a clear filter or without any filters were
adjusted to the $V$-band observation by TRISPEC/KANATA by adding
constants.   

\section{Results}

\subsection{Overall behavior of the 2006 outbursts of J1021}

\begin{figure}
  \begin{center}
    \FigureFile(80mm,80mm){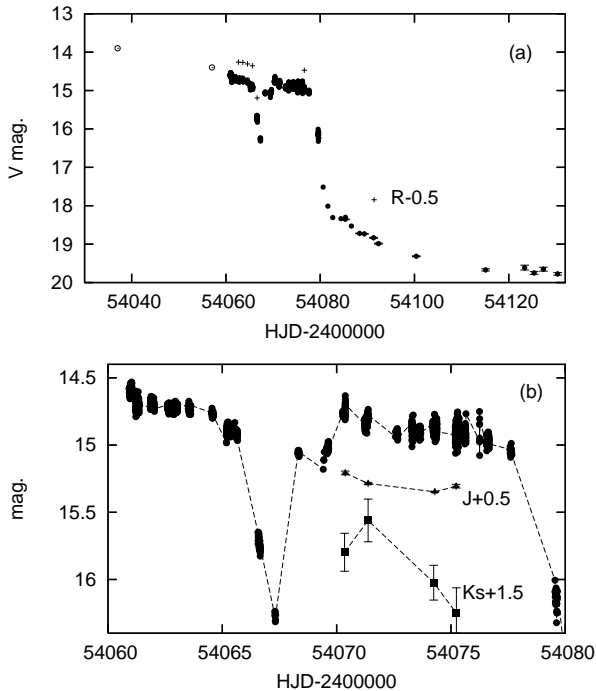}
    %%% \FigureFile(width,height){filename}
  \end{center}
  \caption{Light curves of the outburst of SDSS~J102146.44$+$234926.3
  in 2006.  The abscissa and ordinate denote the time in HJD and the
  magnitude, respectively.  (a)~Whole light curve of the
  outburst.  The filled circles are our $V$-band and unfiltered CCD
  observations.  The crosses are one-day averages of $R_{\rm
  c}$-magnitudes, which are shifted by adding $-0.5$~mag, as indicated
  in the figure.  The open circles represent the $V$-band observations 
  reported in \citet{chr06cbet746}.  (b)~Enlarged light curve of
  the main outburst, the dip, and the rebrightening phase.  The filled
  triangles and squares are $J$- and $K_{\rm s}$-bands observations,
  respectively.  These infrared points are averaged magnitudes of
  each night.  The magnitudes of the infrared points are shifted
  in this figure by adding constants of 0.5~mag ($J$) and 1.5~mag
  ($K_{\rm s}$), as indicated in the figure.}\label{fig:lc}
\end{figure}

According to \citet{chr06cbet746}, J1021 had already be in outburst on
JD~2454037 with $V=13.9$.  We confirmed the outburst on JD~2454060,
23~d after the first record of the outburst.  Our observation campaign
then started, revealing the overall behavior of the outburst and a
subsequent fading phase, as shown in figure~\ref{fig:lc}.  The
panel~(a) includes all observations from the discovery to a fading
phase from the outburst.  The event can be divided in 5 phases, that
is, the main outburst (JD~2454037---2454065), the dip
(JD~2454066---2454067), the rebrightening (JD~2454068---2454077), the
rapid fading phase (JD~2454079---2454081), and the long fading tail
(JD~2454082---).

The total duration from the main outburst to the rebrightening was at
least 40~d.  The average fading rate was 0.05~mag~$\,{\rm d}^{-1}$
during the main outburst.  Compared with typical durations of
superoutbursts (10---15~d; \cite{war95book}), the total duration is
atypically long.  It is rather reminiscent of WZ~Sge stars whose total 
duration of outbursts (including rebrightening phases) is 60---100~d
(\cite{kat01hvvir}).

The main outburst was terminated by the dip, whose duration is
$\lesssim 3\,{\rm d}$.  The object, then, experienced the
rebrightening.  The panel~(b) of figure~\ref{fig:lc} shows light
curves during the main outburst, the dip, and the rebrightening phase.
The rebrightening continued for 10---11~d.  These features of the
light curve is quite analogous to the dip---long-lived plateau
structure observed in several WZ~Sge stars, that is, AL~Com, CG~CMa,
TSS~J022216.4$+$412259.9 and V2176~Cyg (\cite{nog97alcom};
\cite{ish02wzsgeletter}; \cite{kat99cgcma}; \cite{ima06j0222};
\cite{nov01v2176cyg}).

The object entered the rapid fading phase from the rebrightening
on JD~2454079.  The rapid fading stopped at 2.4~mag brighter than its 
quiescence.  The object, then, exhibited the long fading tail.  It
takes at least 50~d until returning to its quiescent level.  This
long fading is also one of characteristics observed in WZ~Sge stars
(\cite{kat04egcnc}).  Ordinary SU~UMa stars returned to their
quiescent level within a few days after superoutbursts.

\subsection{Rising phase of the rebrightening}

We found that the rising trend to the rebrightening maximum was not
monotonous.  The rebrightening was first observed on JD~2454068.  With
the observation on JD~2454067, the rising rate is calculated to be
$<-1.2\,{\rm mag}\,{\rm d}^{-1}$.  The rising trend apparently became
weak or stopped between JD~2454068---2454069, as can be seen from
the figure~\ref{fig:lc}.  We detected a clear brightening trend,
$-0.47\pm 0.06\,{\rm mag}\,{\rm d}^{-1}$, in the data on JD~2454069.
The object, then, reached the rebrightening maximum on JD~2454070.  
The rapid rising trend was, thus, temporarily terminated on
JD~2454068 and restarted on JD~2454069.  

Similar structures in light curves were also observed in other
WZ~Sge stars showing long plateau-type rebrightenings.  In the case of
the 1995 outburst of AL~Com, a rapid rising from the dip was followed
by a temporary fading.  As a result, a clear precursor appeared just
before the rebrightening maximum (\cite{nog97alcom}).  In the case of
V2176~Cyg, we can see a clear fading trend in the first day of a
rebrightening (\cite{nov01v2176cyg}).  This feature also indicates the 
presence of a precursor before the rebrightening maximum.

The observed structure in J1021 can also be interpreted as a sign
of a precursor before the rebrightening maximum.  These precursors may
be a common characteristic in long plateau-type rebrightenings in
WZ~Sge stars.  The durations of the precursors are $< 2\,{\rm d}$ in
all cases, which also supports the scenario that the observed
precursors have the same nature.

\subsection{Color variations}

\begin{figure}
  \begin{center}
    \FigureFile(80mm,80mm){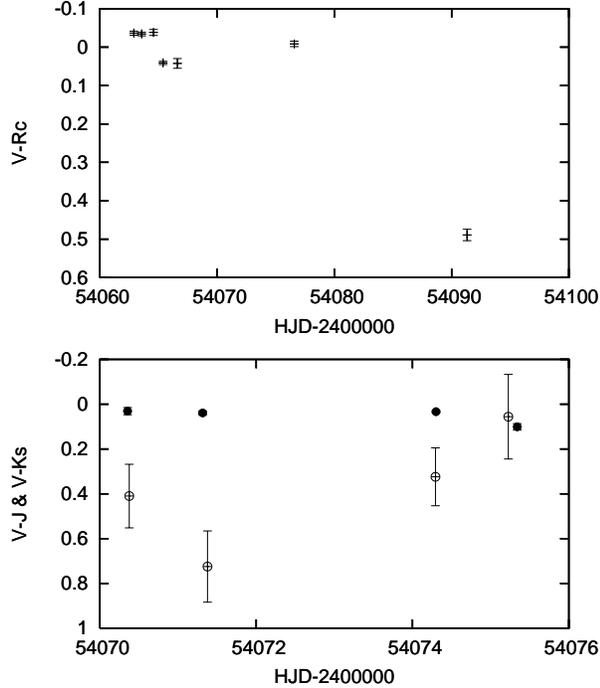}
    %%% \FigureFile(width,height){filename}
  \end{center}
  \caption{Temporal color variations.  The abscissa and ordinate
    denote the time in HJD and the color indices.  Upper panel:
    Variations of $V-R_{\rm c}$.  Lower panel: Variations of $V-J$ and
    $J-K_{\rm s}$.  The filled and opened circles are the $V-J$ and
    $J-K_{\rm s}$, respectively.}\label{fig:color} 
\end{figure}

We performed simultaneous $V$- and $R_{\rm c}$-band observations, as
shown in the panel~(a) of figure~\ref{fig:lc}.  Using these data, we
show temporal variations of the color $V-R_{\rm c}$ in the upper panel
of figure~\ref{fig:color}.  The object first remained at $V-R_{\rm
  c}=-0.04$ during the main outburst, then slightly reddened to 
$V-R_{\rm c}=+0.04$ just before the rapid fading stage.  During the
rebrightening phase, it again became bluer ($V-R_{\rm c}=-0.01$).  A
quite red color was observed during the long fading tail 
($V-R_{\rm c}=+0.49$). 

Our observations revealed the IR behavior of WZ~Sge stars during a
rebrightening phase for the first time.  The $J$- and $K_{\rm s}$-band  
light curves are shown in the panel~(b) of figure~\ref{fig:lc}.  The
$J$-band light curve follows a temporal evolution similar to the
optical one, that is, a gradual fading trend since the rebrightening
maximum and a slightly brightening at $\sim$JD~2454075.  The $K_{\rm
  s}$-band light curve, in contrast to the $J$-band one, the maximum
apparently delayed to the optical one, which is followed by a rather
rapid fading.  As a result, the color variations also show different
behaviors between $V-J$ and $J-K_{\rm s}$, as can be seen in the lower
panel of figure~\ref{fig:color}.  As reported in \textsection~3.4 and
3.5, there are periodic short-term humps in light curves.  The colors
in figure~\ref{fig:color} were calculated from the magnitudes at the
bottom of the humps in order to avoid the contribution from the
variable hump component.

As can be seen in figure~\ref{fig:color}, the color index of $V-J$
almost remains constant at $\sim 0.03$ in JD~2454070---2454075.  The
color $J-K_{\rm s}$, on the other hand, exhibits dramatic variations
from $J-K_{\rm s}=0.7$ to $0.1$.  The $V-J$ color during the
rebrightening is typical for the color of dwarf novae in outburst
($B-V\sim 0$; \cite{bai80DNcolor}).

Spectra during dwarf nova outbursts are described with the
multi-temperature blackbody model (, or the standard disk model;
\cite{sha73disk}).  The temperatures of the disk in dwarf nova
outbursts are from $\sim 10^5\,{\rm K}$ to $\sim 10^4\,{\rm K}$ at the
innermost and outermost region, respectively (e.g., \cite{hor85zcha}).
The emission between the $V$- and $J$-bands, hence, originates from
the outermost region of the disk.   As a result, this region of
spectra can be well described with a simple blackbody spectrum.  We
corrected the interstellar reddening for the observed color $V-J$
using the maximum reddening for the direction of J1021 expected to be
$E(V-J)=0.05$ based on the database in \citet{sch98reddening}.  In the
case of J1021 in the rebrightening phase, the intrinsic color
$V-J=0.08$ indicates a color temperature of 9000~K.  The blackbody
approximation is also supported by the blue color of $V-R_{\rm
  c}=-0.01$ during the rebrightening phase.  

\begin{figure}
  \begin{center}
    \FigureFile(80mm,80mm){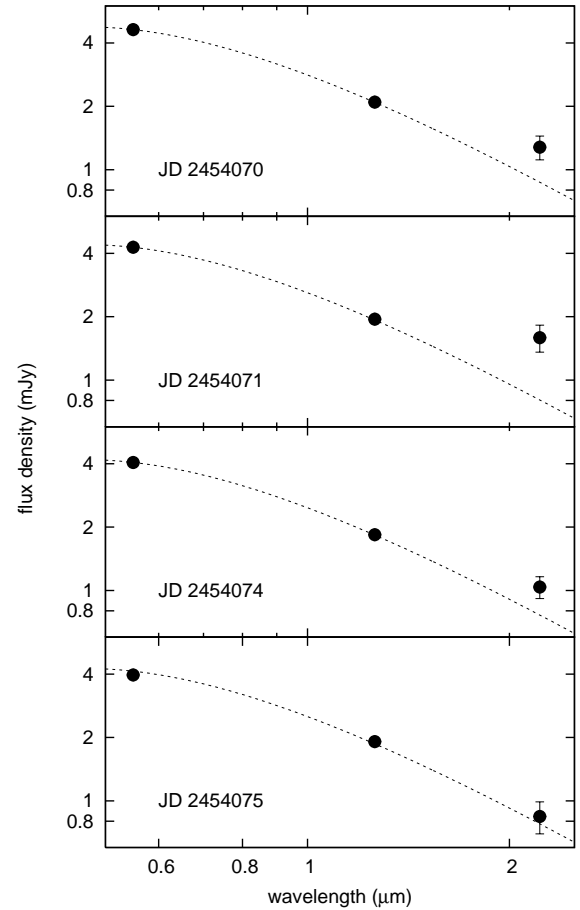}
    %%% \FigureFile(width,height){filename}
  \end{center}
  \caption{Spectral energy distributions between $V$- and $K_{\rm
 s}$-bands during the rebrightening.  The abscissa and ordinate denote
 the wavelength in ${\rm \mu m}$ and the flux density in mJy,
 respectively.  The figure has 4 panels for each observation date,
 that is, JD~2454070, 2454071, 2454074, and 2454075.  The filled
 circles are our $V$-, $J$-, and $K_{\rm s}$-band observations.  The
 dashed lines indicate a blackbody spectrum with a temperature of
 9000~K.}\label{fig:sed}
\end{figure}

Figure~\ref{fig:sed} shows the temporal development of spectral energy 
distributions (SED) during the rebrightening phase.  The filled
circles are our $V$-, $J$-, and $K_{\rm s}$-band observations.  The
dashed lines indicate a blackbody spectrum with a temperature of
$9000\,{\rm K}$.  As can be seen in the figure, there are significant
excesses in the $K_{\rm s}$-band flux over the blackbody spectrum.
The excess reached the maximum at JD~2454071, and then, rapidly
decreased with time.  In JD~2454075, there is no significant excess in
the $K_{\rm s}$-band, and the SED can be described with the simple
blackbody in the $V$-, $J$-, and $K_{\rm s}$-bands regime.  These
results strongly indicate the appearance of another emission source
outside the high temperature disk.  This component was dominant in
wavelengths longer than $\sim 2\,{\rm \mu m}$ for the first couple of
days of the rebrightening, and then, disappeared in a time-scale of a
few days.
  
\subsection{Temporal evolution of optical superhumps}

\begin{figure}
  \begin{center}
    \FigureFile(80mm,80mm){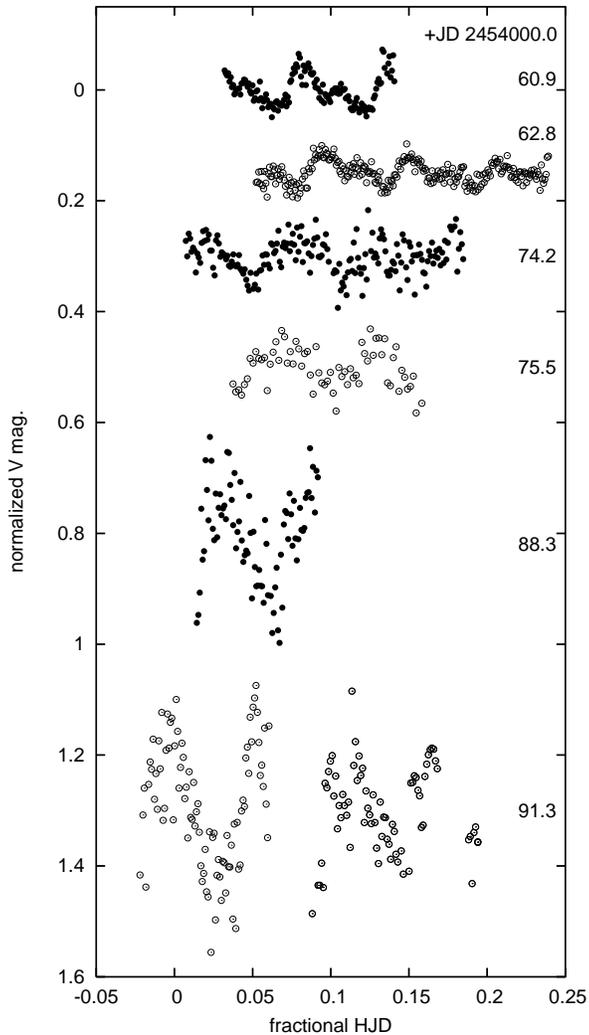}
    %%% \FigureFile(width,height){filename}
  \end{center}
  \caption{Examples of short-term variations during the main outburst
    (JD~2454060 and 2454062), the rebrightening phase
    (JD~2454074 and 2454075), and the fading tail (JD~2454088 and
    2454091).  The abscissa and ordinate denote the fractional HJD and
    the $V$-band magnitudes, respectively.  The magnitudes are
    normalized by their averages and shifted by constants.}\label{fig:lcs}
\end{figure}

We detected short-term periodic modulations during the outburst.
Examples of them are shown in figure~\ref{fig:lcs}.  They have common 
characteristics of $\lesssim 0.1\,{\rm mag}$ amplitudes, except for
those observed during the early fading tail which have larger
amplitudes of 0.3---0.4~mag.  During the main outburst, they have
clear sawtooth profiles, which are typical for superhumps in
SU~UMa-type dwarf novae in superoutburst.  

\begin{figure}
  \begin{center}
    \FigureFile(80mm,80mm){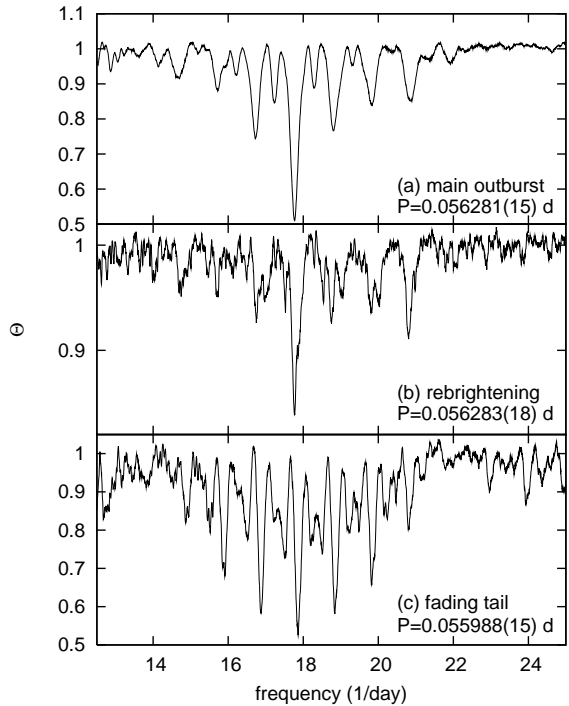}
    %%% \FigureFile(width,height){filename}
  \end{center}
  \caption{Frequency---$\Theta$ diagrams of superhumps in J1021.  The
 abscissa and ordinate denote the frequency in ${\rm d}^{-1}$ and
 the $\Theta$ defined in the PDM method (\cite{PDM}).  The panels of
 (a), (b), and (c) present the results calculated by the data in the
 main outburst, the rebrightening phase, and the fading tail,
 respectively.}\label{fig:pdm}
\end{figure}

We calculate periods of the modulations in the main outburst, the
rebrightening phase, and the fading tail.  We performed period
analysis using the Phase Dispersion Minimization (PDM) method
(\cite{PDM}) for the data during the main outburst
(JD~2454060---2454065), the rebrightening phase
(JD~2454068---2454077), and the fading tail (JD~2454082---2454091).
Note that our sample of the fading tail only covers its early phase.
Using those data, we calculated the $\Theta$ defined in the PDM method
for each frequency.  The obtained frequency-$\Theta$ diagrams are shown in
figure~\ref{fig:pdm}.  The best periods are calculated to be $P_{\rm
main}=0.056281\pm 0.000015\,{\rm d}$, $P_{\rm rebr}=0.056283\pm
0.000018\,{\rm d}$, and $P_{\rm tail}=0.055988\pm 0.000015\,{\rm d}$
for the main outburst, the rebrightening, and the fading tail,
respectively.  The period in the main outburst is in agreement with
that in the rebrightening within the errors, while $P_{\rm tail}$ is
significantly shorter than those periods.  These short periods are
typical for superhump periods in WZ~Sge stars.  In conjunction with
the appearance of the long-lived rebrightening and the fading tail, we 
conclude that J1021 is a new member of WZ~Sge stars (also see,
\cite{gol07j1021}).  The superhump period of J1021 is, hence, $P_{\rm
  SH}$=$P_{\rm main}$=$P_{\rm rebr}$.

In the case of J1021, we can find no sign of early superhumps which
are known as the most convincing evidence for the WZ~Sge nature
(\cite{kat04egcnc}).  The lack of early superhumps is naturally
explained because they appear only in the first week of the main
outburst, in which we had no time-series data in J1021. 

\begin{figure}
  \begin{center}
    \FigureFile(80mm,80mm){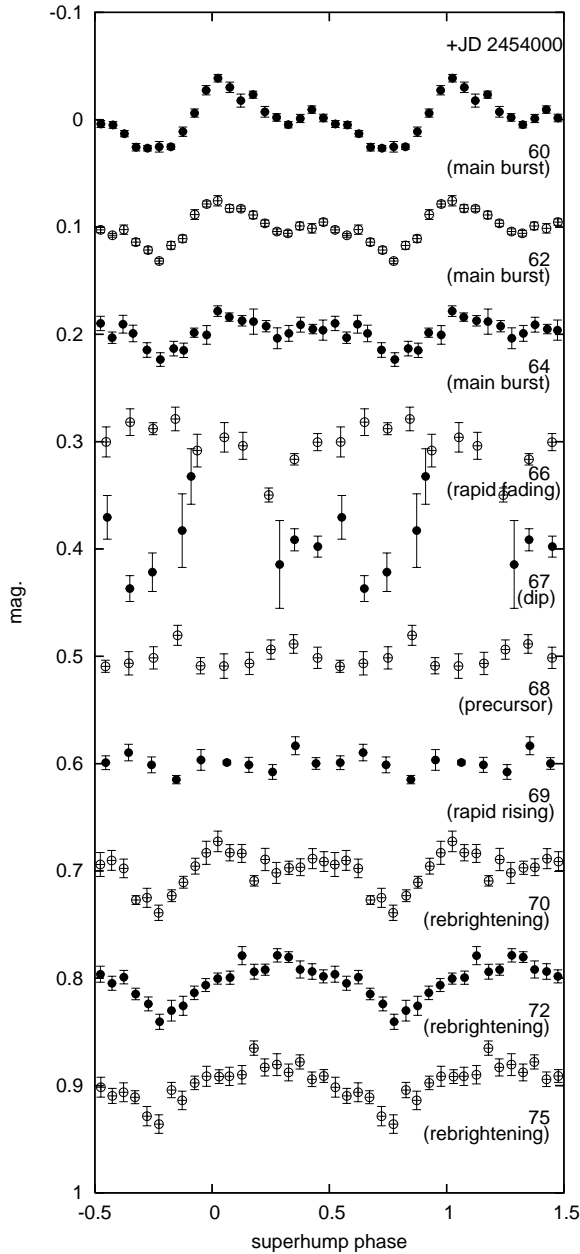}
    %%% \FigureFile(width,height){filename}
  \end{center} 
  \caption{Phase-averaged light curves of superhumps.  The abscissa
 and ordinate denote the superhump phase and the differential
 magnitude, respectively.  The superhump phase is defined by $P_{\rm
 SH}=0.056281$ and an arbitrary epoch.  The light curves are 
 normalized by their average magnitudes for each night and shifted to
 be readily compared.  Observation dates and states during the
 outburst are also indicated in the figure.}\label{fig:hump}
\end{figure}

Figure~\ref{fig:hump} presents the temporal evolution of superhumps.
The light curves are folded with $P_{\rm SH}$ for each night.  As can
be seen from the figure, the superhump amplitude decreased with time
during the main outburst (JD~2454060---2454064, labeled ``60'',
``62'', and ``64'' in the figure).  The superhump profiles consist of
the main hump at the phase $\sim 0$ and the secondary hump at the
phase $\sim 0.5$, which is a typical characteristic for the late phase
of superoutbursts in ordinary SU~UMa stars (\cite{pat03SH}).  During
the rapid fading phase from the main outburst, the phase of humps was
apparently inverted (``66''), while this trend is not confirmed in
later observations (``67'').  We can only find a sign of modulations
during the dip.  The superhump signal, then, weakened during the
precursor and just before the precursor maximum (``68'' and ``69'').
The regrowth of superhumps was observed at the precursor maximum.
Their profile and phase are quite analogous to those observed during
the main outburst, having the main and secondary humps (``70'').  The
profile changed to a flat-top one in the late rebrightening phase,
while their overall phase apparently unchanged (``72'' and ``75'').  

\begin{figure}
  \begin{center}
    \FigureFile(80mm,80mm){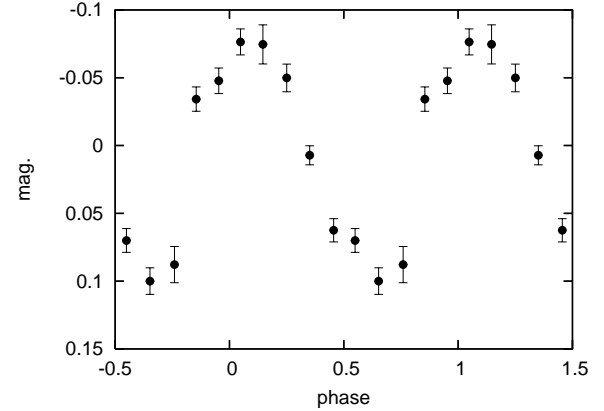}
    %%% \FigureFile(width,height){filename}
  \end{center}
  \caption{Phase-averaged light curves of the humps observed during
 the fading tail.  The symbols are same as in figure~\ref{fig:hump},
 except for the hump period for folding the light curve (0.055988~d).}
  \label{fig:late}
\end{figure}

As mentioned above, a significantly shorter periodicity was found in 
the early fading tail.  The averaged profile of the short-term
modulations is shown in figure~\ref{fig:late}.  The most noteworthy
feature of the modulations is a large amplitude of $\sim 0.2$~mag
compared with those in the superhumps during the superoutburst.  

\begin{figure}
  \begin{center}
    \FigureFile(80mm,80mm){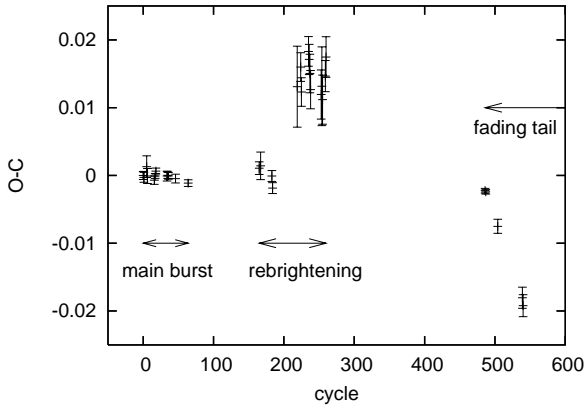}
    %%% \FigureFile(width,height){filename}
  \end{center}
  \caption{$O-C$ diagram of superhumps in J1021.  The abscissa and
  ordinate denote the cycle of superhumps and the $O-C$ in days,
  respectively.  The corresponding phases are also indicated in the
  figure.}\label{fig:o-c}
\end{figure}

It is well known that superhump periods tend to increase 
throughout superoutbursts of WZ~Sge stars (\cite{kat01hvvir}).  In
order to search possible variations of the superhump period in J1021,
we determined the peak times of observed superhumps.  The peak times
were calculated by taking cross-correlations between observed light
curves and templates of superhump profiles.  The templates are
averaged profiles during the main outburst, the rebrightening phase,
and the fading tail.  Using the determined peak times of the
superhumps, we calculated $O-C$s of them with the period of $P_{\rm
  SH}$ and the epoch of HJD~2454060.979634, the first superhump maxima
we observed.  As a result, we obtained the $O-C$ diagram depicted in
figure~\ref{fig:o-c}.  The superhumps maintained constant $O-C$ until
the rebrightening maximum.  A fitting of a quadratic curve for $O-C$s
between the cycle $0$ and $200$ yielded no significant period
derivative ($\dot{P}/P=2.1\pm 2.5\times 10^{-6}$).  In
figure~\ref{fig:o-c}, the $O-C$ apparently jumped during the
rebrightening phase.  As can be seen from figure~\ref{fig:hump}, this
is attributed not to the phase shift of overall superhumps, but to the
variation of the superhump profile and the peak phase of the
superhumps.  We confirm the shorter periodicity in the early fading
tail in figure~\ref{fig:o-c}. 

\subsection{Infrared superhumps during the rebrightening phase}

\begin{figure}
  \begin{center}
    \FigureFile(80mm,80mm){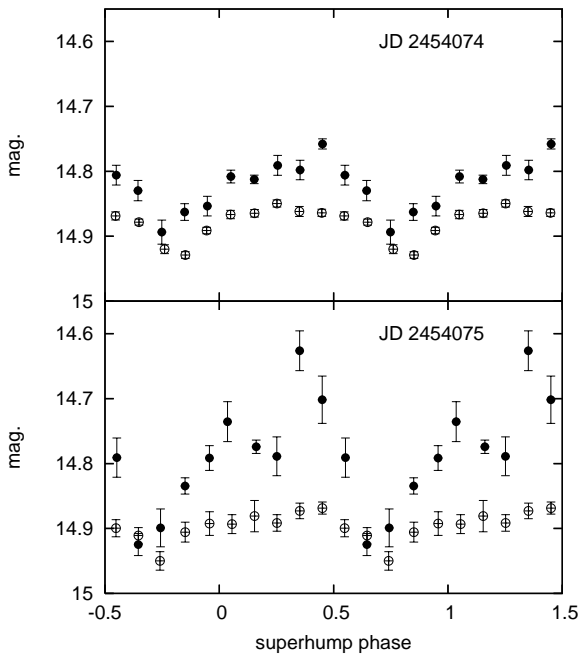}
    %%% \FigureFile(width,height){filename}
  \end{center}
  \caption{Optical and IR superhump profiles during the rebrightening
  phase.  The abscissa and ordinate denote the superhump phase and the
  $V$ and $J$ magnitude, respectively.  The superhump phase is
  defined by the same period and epoch as
  in \ref{fig:hump}.  The filled and open circles represent the $J$-
  and $V$-band superhumps, respectively.  The upper and lower panels
  are observations on JD~2454074 and 2454075,
  respectively.}\label{fig:IRhump} 
\end{figure}

We succeeded in obtaining simultaneous $V$- and $J$-band light curves of
superhumps on JD~2454074 and 2454075 during the rebrightening phase.
Figure~\ref{fig:IRhump} presents superhump
profiles in the $V$- and $J$-bands.  The superhump amplitude in the
$J$-band is significantly larger than that in the $V$-band.  Between
JD~2454074 and 2454075, the $J$-band superhumps, furthermore, evolved
in amplitude.  A clear double-peak profile is seen in the superhumps
on JD~2454075, and a hint of the same profile can also be seen on
JD~2454074.  The phases of these double peaks are apparently in
agreement with those of the optical superhumps observed during the
main outburst and the early phase of the rebrightening, as can be seen 
from figure~\ref{fig:hump}.  These unprecedented ``IR superhumps''
indicate that the superhump source releases its energy not mainly in
the optical, but in the IR regimes. 

In previous multicolor optical observations, the color at the
superhump maxima is only slightly redder than that at the bottom of
the superhumps (\cite{sto84tumen}; \cite{has85ektra};
\cite{bru96oycareclipsemapping}).  The superhump temperature has been
estimated to be 6000---10000~K, which is considered as the temperature
at the outermost region of the accretion disk in outburst.  On the
other hand, \citet{sma05SH} has recently suggested that a higher
temperature ($\gtrsim 15000\,{\rm K}$) is theoretically required for
superhump sources.  

In order to estimate the temperature of the IR superhump source in
J1021, we consider two components of blackbody emissions, that is, a
variable (superhump) component and an invariable inner disk component.
We calculate the color temperatures for each components.  The color 
temperature of the invariable component is determined from the $V-J$
color at the bottom of the superhumps.  Using this color temperature,
we can calculate the color temperature and the size of the emitting
area of the variable component from the V- and J-band amplitudes of
the superhumps.  The size of the emitting area of the variable
component is a relative one to that of the invariable component.  The
interstellar reddening was corrected as mentioned above.

The results are summarized in table~\ref{tab:bbsim}.  The temperature
of the invariable component is $9000\,{\rm K}$ in both dates,
which is typical for the outermost region of accretion disks in
outburst.  The temperature of the variable component is, on the other
hand, much lower than that of the invariable component.  The evolution 
of the IR superhumps between JD~2454074 and 2454075 requires both a
lower temperature and a larger emitting area.  The temperature of
4600~K on JD~2454075 is atypically low compared with superhump
temperatures previously reported in SU~UMa stars.  The large emitting 
area of the variable component on JD~2454075 is, in particular,
remarkable.  In conjunction with the constant temperature and the
magnitude of the invariant component, the evolution of the IR
superhumps implies a rapid outward expansion of the IR superhump
source from JD~2454074 to 2454075.  In addition, the large relative
emitting area of IR superhumps indicates that the actual size of the 
inner, invariable hot disk is small.   

\begin{table}

\caption{Color temperatures of variable and invariable components.}
\label{tab:bbsim}
\begin{center}
\begin{tabular}{cccc}
\hline \hline
JD & $T_{\rm base}$ & $T_{\rm var}$ & Emitting area$^*$ \\
\hline
2454074 & $9000\pm 300$ & $6400\pm 400$ & $0.3\pm 0.2$ \\
2454075 & $9000\pm 300$ & $4600\pm 300$ & $1.4\pm 0.4$ \\
\hline
\multicolumn{4}{l}{\footnotesize * The emitting area of the variable component}\\
\multicolumn{4}{l}{\footnotesize relative to that of the invariable one}\\
\end{tabular}
\end{center}
\end{table}

\subsection{Active low temperature region at the outermost part of
  accretion disks}

As mentioned above, our observations of J1021 revealed unprecedented
IR activities during the rebrightening phase, namely the $K_{\rm
  s}$-band excess (\textsection~3.3) and the IR superhumps
(\textsection~3.5).  These results strongly indicate the presence of a
cool component outside the hot inner disk during the rebrightening.
Our detection of the prominent IR superhump, moreover, revealed that 
the outer disk is not only cool, but also active in terms of the tidal
instability.  The tidal dissipation, hence, still enhances the mass
accretion in the outer disk.  The observed IR superhumps require an
extensive cool component over the 3:1 resonance radius, which means
that a substantial amount of matter is stored and active outside the
hot inner disk.

The optical behavior of J1021 during the rebrightening phase is
analogous to those previously observed in other WZ~Sge stars
(e.g., \cite{nog97alcom}).  There is no hint for the cool component
only with the optical data of J1021.  To date, optical observations
during rebrightening phases have failed to find a clue for a mass
reservoir outside the hot inner disk, while the presence of it has
been expected (\cite{kat98super}; \cite{hel01eruma}; \cite{osa01egcnc};
\cite{kat04egcnc}).  The IR active region of J1021 finally provides 
evidence for the presence of the mass reservoir during the
rebrightening phase.  We propose that the strong tidal dissipation
working at the cool active region maintains a moderately high mass
accretion rate at the outermost area, which sustains the hot state of
the inner disk.

\citet{pat98egcnc} reported an unusually red color of $V-I\sim 0.9$
and strong Na~D absorption during the rebrightening phase of a WZ~Sge
star, EG~Cnc.  These features can be retrospectively interpreted as a
hint of the extensive cool component in the outer disk, although the
original authors implied a rather usual quiescent disk behind a
cooling wave.  We note that another hint of the cool component can be
seen during the rebrightening phase of WZ~Sge.  According to
\citet{how04wzsge}, at the rebrightening maximum of WZ~Sge, the color
temperatures derived from $B$-, $V$-, and $R$-band observations are
9000---10000~K, although the observed $R-I$ indicates only 7700~K.
This implies an $I$-band excess over the $\sim 10000$~K disk.  This can
also be interpreted with the cool mass-reservoir.  A sign of the cool
component was also observed as deep Na~D absorption lines during the
main outburst of WZ~Sge (\cite{nog04wzsge}). 

In the theoretical framework of dwarf novae, state transitions between
a cool and a hot disk must occur in an entire disk (\cite{sma84DI};
\cite{min85DNDI}).  The temperature of the IR superhumps is, however,
much lower than those permitted for the hot disk ($\gtrsim 8000\,{\rm
  K}$).  Our observation of J1021 during the rebrightening indicates
that it had no entire hot disk, but had a two-component disk, namely a
hot inner disk and a cool or transitional state disk at the outermost
part.  Such a simultaneous presence of hot and cool disks could be 
problematic for the thermal instability model.

\section{Discussion}

\subsection{Infrared behavior during dwarf nova outbursts in general}

In the previous section, we reported the IR activity during the
rebrightening phase of J1021.  The uniqueness of its IR behavior among
SU~UMa stars are, however, unclear due to the lack of enough samples
for IR observations during dwarf nova outbursts.  \citet{nay87oycar}
reported optical and IR observations during a superoutburst of an
ordinary SU~UMa-type dwarf nova, OY~Car.  In this paper, optical and
IR superhumps have similar amplitudes, that is, $\sim 0.3$~mag both in
$B$- and $J$-bands.  They are consistent with the classical picture of
the accretion disk in dwarf nova outbursts, in which a superhump
source having 8000---10000~K lies at the outermost part of the disk
having $\gtrsim 10000$~K.  The IR behavior of OY~Car is totally
different from that observed in the rebrightening phase of J1021.  To
date, this observation of OY~Car is all information about IR behavior
during superoutbursts.  About J1021, furthermore, we have no IR data
during the main outburst which should be compared with those during
the rebrightening.  In order to reveal the nature of the IR active
region, future IR observations are required during superoutbursts and
rebrightenings of ordinary SU~UMa and other WZ~Sge stars.

\subsection{Validity of the color temperature}

In the previous section, we calculated the color temperatures of
superhump sources from observed colors.  As a result, we obtained low
color temperatures, which were rather close to those observed at
quiescence.  The color temperature provides a good approximation if
the source is optically thick.  In dwarf novae, however, the optical
light from the accretion disk at quiescence is, in general, dominated
by the optically thin emission (\cite{how04wzsge}, and references
therein).  Even in outbursts, the continuum radiation at the outermost
region is predicted to be optically thin
(\cite{wil80ADemissionline}).  In the case of J1021 during the
rebrightening, the temperature of the IR superhumps was rather close
to the quiescent one, although the entire disk luminosity was still
high.  The radiation mechanism of the IR superhump is, hence,
less obvious.  If the IR emission is optically thin, the color
temperatures of the low temperature region in table~\ref{tab:bbsim}
may have no complete validity as an indicator of real temperatures.
We, then, need to solve the radiation transfer in the disk in order to 
reproduce the observed light curves, which is, however, beyond the
scope of this paper.  Results of such a full calculation, on the other
hand, have only little influence on our order-of-magnitude discussion.
IR spectroscopic observations are required to conclude whether the
outer low temperature region is optically thin or thick during the
rebrightening phase.

\subsection{Delayed maximum of rebrightenings in WZ~Sge stars}

Among 5 systems showing long plateau-type rebrightenings, 3 systems
exhibit a precursor just before their rebrightening maximum, as
mentioned in \textsection~3.2.  In ordinary SU~UMa stars, precursors
are occasionally observed just before supermaxima.  They are
interpreted as a normal outburst triggering a subsequent superoutburst
(\cite{osa03DNoutburst}).  After the precursor maximum, the object
starts a rapid fading since it takes a few days for the full evolution
of the tidal instability.  When the tidal dissipation begins working
effectively, the rapid fading is terminated by a rising trend toward
the supermaxima.  The similar mechanism may also work in the structure
of the long plateau-type rebrightening.

Light curves between precursors and rebrightening maxima are, however,
peculiar in WZ~Sge stars compared with the precursor---supermaximum
structure in ordinary SU~UMa stars.  Several WZ~Sge stars keep a
gradually rising trend after initial rapid risings of rebrightenings.
AL~Com and V2176~Cyg, for example, reached the maximum of
rebrightenings $\sim 5$ and $\sim 4\,{\rm d}$ after the precursors,
respectively (\cite{nog97alcom}; \cite{nov01v2176cyg}).
TSS~J022216.4$+$412259.9 is an extreme case, in which a gradual rising
apparently continued for $\sim 8\,{\rm d}$ until the rebrightening
maximum (\cite{ima06j0222}).  This characteristic is clearly different
from that observed in ordinary superoutbursts in which objects reach
their supermaximum within a day after the onset of outbursts
(\cite{war95suuma}).  While J1021 shows no delay of the rebrightening
maximum, we discuss the delayed maximum in terms of the mass reservoir
at the outermost part of the disk.

According to the disk instability model, the rising speed corresponds
to the propagation speed of the heating wave in the hot disk
(\cite{min85DNDI}).  The delayed rebrightening maximum is possibly
attributed to a late propagation of the heating wave into the outer
mass reservoir, namely the IR active region.  If the IR superhump
observed in J1021 is a characteristic feature for the rebrightening
phase, however, this scenario is unlikely because it would yield an
entire hot disk and prominent optical superhumps.

Another possible scenario for the delayed rebrightening maximum is
that the inner hot disk continuously receives a sufficient amount of
gas from the outer mass reservoir.  In the case of J1021, the tidal
dissipation effectively works at the outer low temperature region as
demonstrated by the IR superhumps.  We can, thus, naturally expect a
significant mass accretion from the outer cool to the inner hot part
of the disk.  

\subsection{Origin of the $K_{\rm s}$-band excess}

During the rebrightening phase, the $K_{\rm s}$-band flux exhibited a 
clear excess over the blackbody component which dominated between the
$V$ and $J$ ranges, as mentioned in \textsection~3.3.  A noteworthy
feature of the $K_{\rm s}$-band light curve is a relatively rapid
evolution compared with $V$- and $J$-band ones.  The $V$- and $J$-band
emission is presumably dominated by the optically thick emission from
the disk.  The evolution time-scale of them is a viscous diffusion
time-scale of the hot optically thick disk.  The time-scale becomes
longer in an outer, in other words, lower temperature region of the
disk.  The rapid evolution of the $K_{\rm s}$-band emission is,
thereby, inconsistent with the theory of optically thick disk.  It
implies that the $K_{\rm s}$-band emission is not from an optically
thick disk, but an optically thin region which is located outside the
optically thick disk.  The source of the $K_{\rm s}$-band emission
might be expelled from the optically thick disk at the onset of the
rebrightening as a result of a rapid expansion of the hot disk. 

\subsection{On the nature of the periodicity in the early fading tail}

The periodic modulations during the early fading tail has a 
slightly shorter period than the superhumps, as can be seen from
figures~\ref{fig:pdm} and \ref{fig:o-c}.  The nature of the shorter
period modulations is unclear.  The period is possibly the orbital
period of J1021.  If it is the case, the superhump excess 
$\varepsilon = (P_{\rm SH}-P_{\rm orb})/P_{\rm orb}$ is calculated to
be $0.0052\pm 0.0003$.  The empirical relationship between the
$\varepsilon$ and the mass ratio of binary systems reported in
\citet{pat01SH} yields the mass ratio of $M_2/M_1=0.024$ for J1021.
The $\varepsilon$ and the mass ratio are the smallest one among
previously known dwarf novae (\cite{pat01SH}).  J1021 is possibly the
most evolved dwarf novae if the period of the modulations in the
fading tail indicate the orbital period.  An alternative possibility
is that they are late superhumps.  Late superhumps are periodic
modulations appearing just after superoutbursts (\cite{woe88vwhyi};
\cite{rol01iyumatomography}).  Their period is in agreement with a
superhump period, while their profile is $\sim 0.5$ phase shifted from
that of superhumps.  In the case of J1021, we cannot confirm a clear
phase shift in figure~\ref{fig:o-c}.  We, however, consider that the
late superhump scenario is more plausible because the long fading tail
indicates an effective tidal dissipation, in which we can expect a
long-lived eccentric disk and the superhump periodicity.  As can be
seen in figure~\ref{fig:o-c}, $O-C$s of humps are apparently changed
smoothly from the main outburst to the rebrightening and the early
fading tail.  This also supports the late superhump scenario.

\section{Summary}

Our observation of J1021 detected superhumps with a period of
$0.056281\pm 0.000015\,{\rm d}$.  The object experienced a long
plateau-type rebrightening after the main superoutburst, which is
observed only in WZ~Sge-type dwarf novae.  In conjunction with the
short superhump period, we conclude that J1021 is a new member of
WZ~Sge stars.  We, furthermore, revealed IR behaviors for the first
time during rebrightening phases of WZ~Sge stars.  J1021 exhibited
unprecedented IR activities, namely the $K_{\rm s}$-band excess and
the IR superhump.  The IR superhump, in particular, indicates the
presence of a substantial amount of remnant matter at the outermost
region of the accretion disk even after the main superoutburst.  We
propose that this outer low temperature region is responsible for
maintaining a hot state of the inner disk during the rebrightening.

The authors are grateful to Dr. Y. Osaki and S. Mineshige, for useful
comments on this research.  We really appreciate the development of
TRISPEC by School of Science, Nagoya University.  This work was partly
supported by a Grand-in-Aid from the Ministry of Education, Culture,
Sports, Science, and Technology of Japan (17684004, 17340054,
18740153, 14079206, 18840032) and by the grant of the Ukrainian Fund
of the Fundamental Research F 25.2/139 and CosmoMicroPhysics program of
the National Academy of Sciences and National Space Agency of Ukraine.
Part of this work is supported by a Research Fellowship of Japan
Society for the Promotion of Science for Young Scientists.  

%%%
% See the manual for the detail.
%%%
%\bibliographystyle{pasjtest1}
%\bibliography{cvs}

\end{document}